\documentclass[12pt]{iopart}

\usepackage{latexsym}
\usepackage{graphicx}
\usepackage{iopams}  

\begin{document}
\title[Tests of mode coupling theory in a simple model for miscible polymer blends]{Tests of mode coupling theory in a simple model for two-component miscible polymer blends}
\author{A. J. Moreno$^{1}$ \footnote[3]{To whom correspondence should be 
addressed (wabmosea@ehu.es)}, J. Colmenero$^{1,2,3}$ }

\address{$^{1}$ Centro de F\'{\i}sica de Materiales (CSIC-UPV/EHU), 
Apartado 1072, 20080 San Sebasti\'an, Spain}

\address{$^{2}$ Departamento de F\'{\i}sica de Materiales, Universidad del Pa\'{\i}s Vasco (UPV/EHU),
Apartado 1072, 20080 San Sebasti\'{a}n, Spain}

\address{$^{3}$ Donostia International Physics Center, Paseo Manuel de Lardizabal 4,
20018 San Sebasti\'{a}n, Spain}

\begin{abstract}
We present molecular dynamics simulations on the structural relaxation of a simple 
bead-spring model for polymer blends.
The introduction of a different monomer size induces a large time scale separation
for the dynamics of the two components. Simulation results for a large set of observables
probing density correlations, Rouse modes, and orientations of bond and chain end-to-end vectors,
are analyzed within the framework of the Mode Coupling Theory (MCT).
An unusually large value of the exponent parameter is obtained.
This feature suggests the possibility of an underlying higher-order MCT scenario
for dynamic arrest.
\end{abstract} 
\pacs{64.70.Pf, 83.80.Tc, 83.10.Rs}


\section{Introduction}

Polymer blends are soft-matter systems which exhibit `dynamic asymmetry' in the meaning
that, starting from two homopolymers with different mobilities,
two separated segmental dynamics can still be observed in the blend.
Phenomenological approaches usually consider thermally driven concentration
fluctuations \cite{kumar} and self-concentration effects induced 
by chain connectivity \cite{lodge} as key ingredients for structural 
relaxation in polymer blends \cite{kumar2}. A recent approach combines self-concentration
effects with ideas of the Adam-Gibbs theory \cite{blendsag}.
For most of the investigated systems, dynamics of the two components
in the blend display qualitatively similar features. 
However, recent experimental results by nuclear magnetic resonance (NMR) \cite{nmr1,nmr2},
dielectric spectroscopy \cite{dielec1,dielec2,dielec3,dielec4,dielec5}, or neutron scattering \cite{genix,tyagi} 
suggest that a rather different scenario arises
when the two homopolymers exhibit very different glass transition temperatures.
Hence, for dilute concentrations of the fast component, 
the two components in the blend exhibit strong dynamic inmiscibility.
A large separation in their relaxation times is observed, which can be of even
12 orders of magnitude in blends of poly(ethylene oxide)/poly(methyl methacrylate) 
(PEO/PMMA) for extreme dilution of PEO \cite{nmr2}. 
In such conditions the motion of the chains of the fast component takes place in 
a slowly relaxing matrix formed by the slow component, providing a connection 
with the problem of confinement in host media with interconnected voids.

We have recently performed an investigation on the structural relaxation dynamics of
a simple bead-spring model for polymer blends \cite{blendpaper}. The introduction of monomer 
size disparity between the two components induces a large time scale separation
for low concentrations of the fast component, which displays unusual
relaxation features. Hence, density-density correlators exhibit logarithmic decays
over time intervals of even four decades and a concave-to-convex
crossover by varying the thermodynamic state point (i.e., the control parameters)
or the wave vector \cite{blendpaper}.
Dynamic features observed for this simplified model are supported 
by recent fully atomistic simulations on the blend PEO/PMMA \cite{genix}.
 
We have discussed the unusual features reported in Ref. \cite{blendpaper} within the framework
of the Mode Coupling Theory (MCT) of the glass transition \cite{mctrev1,mctrev2},
and suggested an underlying higher-order MCT transition as the origin of the observed
anomalous relaxation scenario. Higher-order MCT transitions were initially predicted
by schematic models \cite{schematic}, and later derived for simplified models
of short-ranged attractive colloids \cite{dawson,sperl}.
These systems show two different mechanisms for dynamic arrest:
steric repulsion characteristic of colloidal systems, and formation of reversible
bonds, induced by the short-ranged attraction. Coexistence of both mechanisms
of very different localization lengths \cite{dawson,sperl} yields a higher-order MCT transition
in a certain region of the temperature-density plane. The mentioned anomalous 
relaxation features are derived from the MCT equations as specific solutions
associated to the higher-order point \cite{dawson,sperl,prlsimA4,zaccarelli1}.

Results for the mean squared displacements and density-density 
correlators in the bead-spring polymer blend of Ref. \cite{blendpaper} 
display striking similarities with qualitative features
associated to higher-order MCT transitions. Similar results have also been observed
in later simulations of binary mixtures of non-bonded particles with large size disparity,
both for soft \cite{mixturepre,mixturejcp} and ultrasoft interactions \cite{mayer}.
Finally, very recent two-dimensional NMR experiments
on a polymer-plasticizer system have revealed logarithmic relaxation for the 
strongly confined plasticizer \cite{bingemann}.
Hence, this collection of similar experimental and simulation results 
suggest a common relaxation scenario for multicomponent systems exhibiting strong dynamic asymmetry.
Moreover, the mentioned analogies with short-ranged attractive colloids
suggest that the higher-order MCT scenario might be a general feature
of systems showing several mechanisms for dynamic arrest.
For the mentioned polymeric and non-polymeric mixtures,
we have suggested bulk-like caging and confinement
\cite{blendpaper,mixturepre,mixturejcp}. These mechanisms would be respectively induced by the presence
of neighbouring small particles and by the slow matrix formed by the large particles.

It is worth mentioning that solutions of the MCT equations for a fluid of hard spheres
confined in a disordered matrix of {\it strictly static} obstacles explicitly reveal the existence
of a higher-order transition \cite{krakoviack}. 
As discussed in Ref. \cite{krakoviack}, the strictly static nature of the matrix
induces differences with the former mixtures, where the matrix shows a slow relaxation.
Hence, though they share common features for the dynamics of the confined component,
a comparison of results between both kind of mixtures must be taken with care.

The test of MCT predictions for the bead-spring blend model reported in Ref. \cite{blendpaper}
was restricted to density-density correlators. In this article we present a systematic
test for a large set of correlators probing different dynamic features
as Rouse modes or orientations of bond and chain end-to-end vectors.
Consistently with MCT predictions, a common set of dynamic exponents provides a good description
of dynamic correlators in the early-middle stage of the structural $\alpha$-relaxation. 
Consistently with previous results \cite{blendpaper}, the unusually large value
obtained for the exponent parameter suggests that the observed
anomalous relaxation features might be associated to an underlying higher-order order MCT scenario.

The article is organized as follows. In Section 2 we summarize the main details of the 
simulated model. In Section 3 we present simulation results for static correlations.
The main predictions of MCT are exposed in Section 4. We discuss within the framework of MCT 
relaxation features of the slow and fast component in, respectively, sections 5 and 6. 
Conclusions are given in Section 7.

\section{Model and simulation details}

The model introduces a binary mixture of bead-spring chains (of the species A and B).
Each chain consists of $N= 10$ monomers of mass $m = 1$. All the monomers in
a same chain belong to the same species (i.e., all them are A-like or B-like).
Two given monomers (placed at a same chain or at different ones) interact through a 
soft-sphere potential plus a quadratic term,
$V_{\alpha\beta}(r) = 4\epsilon[(\sigma_{\alpha\beta}/r)^{12} - C_0 + C_2(r/\sigma_{\alpha\beta})^{2}]$,
where $\epsilon=1$ and $\alpha$, $\beta$ $\in$ \{A, B\}. 
The interaction is zero beyond a cutoff distance $r _{\rm c} = c\sigma_{\alpha\beta}$, with $c = 1.15$.
Continuity of potential and forces at $r = r_{\rm c}$
is guaranteed by setting the values $C_0 = 7c^{-12}$ and $C_2 = 6c^{-14}$.
The radii of the different pair interactions are $\sigma_{\rm BB} =1$, $\sigma_{\rm AA} = 1.6\sigma_{\rm BB}$, 
and $\sigma_{\rm AB}=1.3\sigma_{\rm BB}$.  
Chain connectivity is introduced by a FENE bonding potential \cite{grest},  
$V^{\rm FENE}_{\alpha\alpha}(r) = -kR_0^2 \epsilon\ln[ 1-(R_0\sigma_{\alpha\alpha})^{-2}r^2 ]$,
between consecutive monomers, with $k=15$ and $R_0 = 1.5$. The superposition of $V_{\alpha\beta}(r)$
and $V^{\rm FENE}_{\alpha\alpha}(r)$ provides an effective bonding potential for connected monomers
with a sharp minimum at $r = 0.985\sigma_{\alpha\beta}$, which makes bond crossing impossible.

The blend composition is defined as
$x_{\rm B} = N_{\rm B}/(N_{\rm A}+N_{\rm B})$, where $N_{\alpha}$ is the number of $\alpha$-chains.
All the data presented here correspond to a fixed composition $x_{\rm B} = 0.3$
(we have simulated a mixture of $N_{\rm A} = 210$ and $N_{\rm B} = 90$ chains).
We use a packing fraction $\phi = (\pi/6)L^{-3}[N_A\sigma_{\rm AA}^3 + N_B\sigma_{\rm BB}^3] = 0.53$,
with $L$ the side of the cubic simulation cell. The value $\phi = 0.53$ is comparable
to those used in simulations of slow relaxation in simple liquids \cite{koband,bennemann}.
In the following temperature $T$, distance, wave vector $q$, 
and time $t$  will be given, respectively, in units of $\epsilon/k_B$, $\sigma_{\rm BB}$,
$\sigma_{\rm BB}^{-1}$, and $\sigma_{\rm BB}(m/\epsilon)^{1/2}$. 

The system is prepared by placing the chains randomly in the simulation cell, with a constraint
that avoids monomer core overlapping.
The Newton equations of motion are integrated in the velocity Verlet scheme \cite{frenkel},
with a time  step ranging from $2 \times 10^{-4}$ to $5 \times 10^{-3}$ for, respectively,
the highest and the lowest investigated $T$. 
Standard periodic boundary conditions are used
for calculation of monomer-monomer distances entering in the interactions. 
Computational expense for the latter calculation
is reduced by implementing a standard link-cell method \cite{frenkel}. 
The system is thermalized at the selected temperature by periodic velocity rescaling.
Then the equilibrium run for data acquisition is performed in the
microcanonical ensemble (constant energy). During this run no drift in thermodynamic quantities is observed,
either aging effects in dynamic correlators computed for different time origins.
Statistical averages at a given state point are performed over typically 20-40 independent runs.

\section{Static properties}

\begin{figure}
\begin{center}
\includegraphics[width=0.57\linewidth]{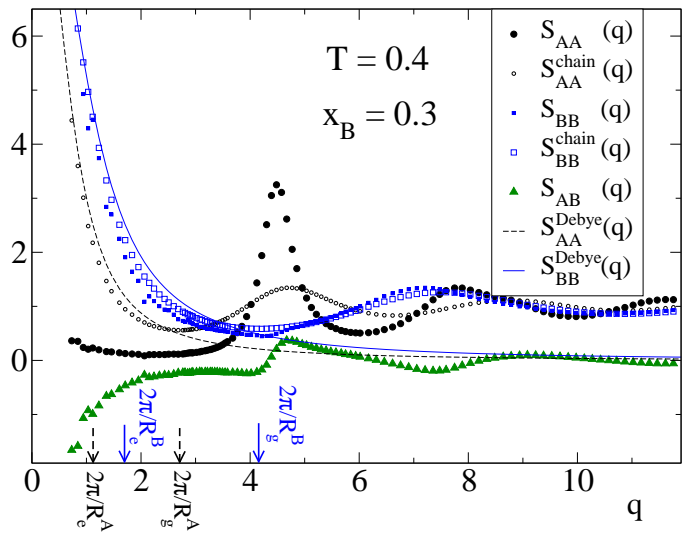}
\end{center}
\caption{Partial static structure factors, $S_{\rm AA}(q)$, $S_{\rm BB}(q)$ and $S_{\rm AB}(q)$,
at $T = 0.4$. Also included are the chain form factors,
$S_{\rm AA}^{\rm chain}(q)$, $S_{\rm BB}^{\rm chain}(q)$, as well as the corresponding Debye functions,
$S_{\rm AA}^{\rm Debye}(q)$, $S_{\rm BB}^{\rm Debye}(q)$. Arrows indicate
the wave vectors $q = 2\pi/R_{\rm e,g}^{\alpha}$, where $R_{\rm e}^{\alpha}$ and $R_{\rm g}^{\alpha}$
are respectively the chain end-to-end distance and gyration radius of the species $\alpha$.}
\label{fig:sq}
\end{figure}

\begin{figure}
\begin{center}
\includegraphics[width=0.51\linewidth]{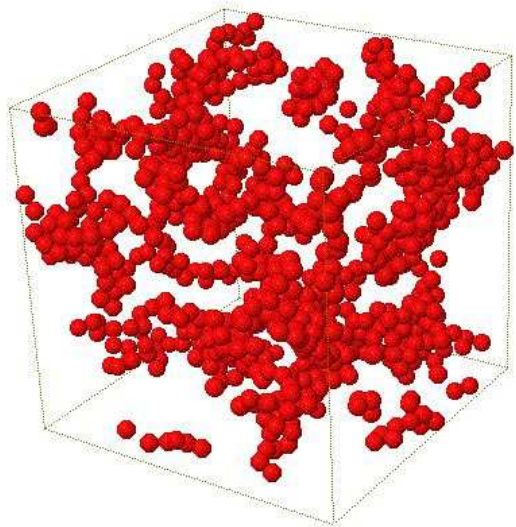}
\end{center}
\caption{A typical configuration of the B-chains.} 
\label{fig:configb}
\end{figure}

In this section we provide information about static correlations in the bead-spring blend.
We compute normalized partial static structure factors, 
$S_{\rm \alpha\beta}(q) = 
\langle \rho_{\alpha}(q,0)\rho_{\beta}^{\ast}(q,0)\rangle /(N \sqrt{N_{\alpha}N_{\beta}})$.
The quantity $\rho_{\alpha}(q,t)$ is the density fluctuation for wave vector $q$ and is defined as
$\rho_{\alpha}(q,t) = \sum_{j}\exp[i{\bf q}\cdot {\bf r}_j^{\alpha}(t)]$,
the sum extending over all the particles of the species $\alpha \in$ \{A,B\}.
Fig. \ref{fig:sq} shows, for a low temperature $T = 0.4$, results for
A-A, B-B, and A-B pairs. Intrachain static structure factors (i.e., chain form factors),
$S_{\rm \alpha\alpha}^{\rm chain}(q)$,
are also displayed. The latter quantities are computed by restricting the product
$\rho_{\alpha}(q,0)\rho_{\alpha}^{\ast}(q,0) = \sum_{j,k}\exp\{i{\bf q}\cdot [{\bf r}_j^{\alpha}(0)-{\bf r}_k^{\alpha}(0)]\}$
over pairs of monomers $j, k$ belonging to a same chain.
A sharp maximum is observed in $S_{\rm AA}(q)$ at $q = 4.5$, which corresponds 
to a typical distance of 1.4 between A-monomers. Results for $S_{\rm AA} (q)$
are qualitatively similar to those reported for the {\it homopolymer} case 
in a similar bead-spring model \cite{bennemann,aichele}. A weak low-$q$ structure
is observed in the present case, which originates from the presence of `holes' in the matrix of A-monomers. 
These holes are created by the inclusion of the B-monomers. 
The negative values of $S_{\rm AB}(q)$ observed at small wave vectors are a signature 
of anticorrelation effects between A- and B-monomers
at large distances, and indicate a moderate degree of demixing. This feature is illustrated
in Fig. \ref{fig:configb}, which shows a typical configuration of the B-chains.
The latter are not homogeneously distributed but form a sort of cluster structure.

The partial static structure factor for B-B pairs, $S_{\rm BB}(q)$, exhibits 
a rather different $q$-dependence (Fig. \ref{fig:sq}). From a comparison with the form factor for B-chains, 
$S_{\rm BB}^{\rm chain}(q)$, it is clear that $S_{\rm BB}(q)$ is largely dominated by intrachain contributions, 
as expected for high dilution of the B-chains in the matrix formed by the A-chains. The peak at $q = 7.2$
corresponds to a typical distance of 0.87 between B-monomers.

Data for the chain form factors in Fig. \ref{fig:sq} are also 
compared with the Debye function \cite{teraoka,doiedwards},
$S^{\rm Debye}_{\rm \alpha\alpha}(q) = 
2Nq^{-4} (R^{\alpha}_{\rm g})^{-4}\{\exp[-q^2 (R^{\alpha}_{\rm g})^2] +q^2 (R^{\alpha}_{\rm g})^2 -1\}$,
which is obtained by assuming a Gaussian distribution of monomer-monomer
distances within the chain \cite{teraoka,doiedwards}. 
As previously observed for the homopolymer case \cite{aichele}, Gaussian statistics approximately
work at low $q$ but clearly break down
for wave vectors probing distances smaller than the chain gyration radius $R^{\alpha}_{\rm g}$. 
The magnitude of the deviations of simulation data from the Debye function is similar to observations
for the homopolymer case \cite{aichele}. Hence, chain statistics is not significantly affected by blending.

\section{Main predictions of MCT}

In this section we summarize some of the main predictions of the Mode Coupling Theory (MCT)
for the glass transition. Extensive reviews can be found, e.g., in Refs. \cite{mctrev1,mctrev2,mctrev3,das,franosch}.
In its ideal version, MCT predicts a sharp transition from an ergodic liquid to a non-ergodic glassy state
for a given value of the relevant control parameter $\xi$ (in the following the temperature, $T$, though
results exposed below are valid for any $\xi$).
On approaching the transition point $T = T_{\rm c}$, MCT establishes a set of quantitative predictions
for any correlator coupled to density fluctuations, $\Phi (t)$. An example is
normalized density-density correlators of wave vector $q$, 
$F_{\alpha\alpha}(q,t) = \langle \rho_{\alpha}(q,t)\rho_{\alpha}^{\ast}(q,0)\rangle /S_{\alpha\alpha}(q)$,
for the species $\alpha$. At the critical temperature $T = T_{\rm c}$ 
the long-time limit of $\Phi (t)$ jumps from zero to a non-zero value, 
denoted as the {\it critical} non-ergodicity parameter, $\Phi^{\rm c}$. 
In the standard case ({\it type-B} transitions) the jump in $\Phi (t)$ is discontinuous, 
i.e., $\Phi^{\rm c}$ takes a finite positive value.

For ergodic states close to the transition point $\Phi$ usually exhibits
a first decay to a plateau, whose time extension increases as the transition is approached.
This plateau regime corresponds to the temporary trapping of each particle within the cage
formed by its neighbouring ones, i.e., the well-known caging effect which is generally present in, e.g., 
supercooled liquids or jammed systems. 
At times longer than the so-called first MCT time scale $t_{\sigma}$, the correlator
$\Phi$ starts a second decay from the plateau to zero. This second decay is commonly
known as the $\alpha$-process and represents the full decorrelation of the system from its initial configuration,
i.e. the structural relaxation.
According to MCT, the initial part of the $\alpha$-process
(denoted as the von Schweidler regime) is given by a power law decay $\propto -t^b$,
with $0 \le b \le 1$. A power-law series expansion extends the description of the $\alpha$-decay to longer times: 
\begin{equation}
\Phi(t) = 
\Phi^{\rm c} -h_\Phi (t/\tau_{\alpha})^{b} + h_\Phi^{(2)}(t/\tau_{\alpha})^{2b} + O(t^{3b}).
\label{eq:vonsch}
\end{equation}
The prefactors $h_\Phi$ and $h_\Phi^{(2)}$ are state point-independent
and are different for each correlator $\Phi$. On the contrary, the von Schweidler 
exponent $b$ is common to all correlators.
The characteristic time scale of the $\alpha$-relaxation, $\tau_{\alpha}$, is the second MCT time scale.
It is also unique for all correlators,
and diverges at the transition point as $\propto (T - T_{\rm c})^{-\gamma}$ (see below). 
The $\alpha$-decay can often be described by an empirical Kohlrausch-Williams-Watt (KWW) function,
$\propto \exp[-(t/\tau)^{\beta_{\Phi}}]$, with a $\Phi$-dependent stretching exponent $0 < \beta_{\Phi} < 1$.
An interesting prediction of MCT \cite{fuchs} is that $\beta_q = b$ and $\tau \propto q^{-1/b}$ 
in the limit of large $q$, both for density-density [$F(q,t)$], and self-correlators, 
$F^{\rm s}_{\alpha}(q,t) = \left\langle \sum_{j} \exp\{i{\bf q}\cdot 
[{\bf r}_{\alpha,j}(t)-{\bf r}_{\alpha,j}(0)]\} \right\rangle/(N N_{\alpha})$.  
This result \cite{mixturejcp,koband,bennemann,aichele,fuchspra,mossa,puertaspre,pisacolme} provides
a consistency test for data analysis.

Another prediction of MCT for state points close to the transition point
is the power law dependence of the diffusivity and the relaxation time $\tau^{\Phi}_{x}$:
\begin{equation}
\tau^{\Phi}_{x}, D^{-1} \propto (T-T_{\rm c})^{- \gamma}.
\label{eq:power}
\end{equation}
The relaxation time $\tau^{\Phi}_{x}$ of the correlator $\Phi$ is defined as the time  
where $\Phi(t)$ decays to some small value $x$, provided it is well below the plateau. 
The time-temperature superposition principle of the MCT establishes that, for $t$ much longer than
the first time scale $t_{\sigma}$,
$\Phi(t/\tau_{\alpha}) = \tilde{\Phi}$, where $\tilde{\Phi}$ is a $\Phi$-dependent scaling function.
According to this prediction, for a fixed $x$ the ratio $\tau^{\Phi}_{x}/\tau_{\alpha}$ 
is temperature-independent for any $\tau^{\Phi}_{x} \gg t_{\sigma}$, i.e, $\tau^{\Phi}_{x} \propto \tau_{\alpha}$.
In other words, $\tau^{\Phi}_{x}$ will be $\Phi$-modulated but will follow the same power law behaviour
in $T$ as the $\alpha$-relaxation time $\tau_{\alpha}$ ({\it even if} $\tau^{\Phi}_{x} \gg \tau_{\alpha}$). 
Note that, in the MCT terminology, $\tau_{\alpha}$
is a single time scale, though its value can be approximately probed by evaluating dynamic correlators $\Phi$
for which $\tau^{\Phi}_{x} \sim \tau_{\alpha}$. This is the case of, e.g., the density-density correlator
$F(q,t)$ for wave vector $q$ at the maximum of the static structure factor $S(q)$, since the former probes
decorrelation over typical distances between nearest-neighbour particles. The relaxation time
of $F(q,t)$ is indeed often denoted as the `$\alpha$-relaxation time', though in the context of MCT
the latter strictly corresponds to $\tau_{\alpha}$.

The exponent $\gamma$ in Eq. (\ref{eq:power}) is given by the relation:
\begin{equation}
\gamma = \frac{1}{2a} + \frac{1}{2b},
\label{eq:gamma}
\end{equation}
with $0 \le a \le 0.395$. Hence $\gamma \ge 1.766$. The critical exponents $a$, $b$, and $\gamma$
are univocally related with the so-called exponent parameter $\lambda$ through:
\begin{equation}
\lambda = \frac{\Gamma^{2}(1+b)}{\Gamma(1+2b)} = \frac{\Gamma^{2}(1-a)}{\Gamma(1-2a)},
\label{eq:lambda}
\end{equation}
where $\Gamma$ is the Gamma function. The exponent parameter $\lambda$ 
is univocally determined by the static correlations (i.e., by the total and partial static structure factors)
at the transition point $T = T_{\rm c}$. For type-B transitions it takes values $1/2 \le \lambda \le 1$. 

When numerical solutions of the MCT equations are not available
the non-ergodicity parameters, prefactors and exponents 
in Eqs. (\ref{eq:vonsch},\ref{eq:power},\ref{eq:gamma},\ref{eq:lambda})
--- which are system-dependent quantities controlled by static correlations ---
are empirically obtained as fit parameters from simulation or experimental data. 
Consistency of the data analysis requires that the so-obtained set of exponents fulfill
both Eqs. (\ref{eq:gamma}) and (\ref{eq:lambda}).

\section{Dynamics of the slow component in the blend \label{secslow}}

\begin{figure}
\begin{center}
\includegraphics[width=0.55\linewidth]{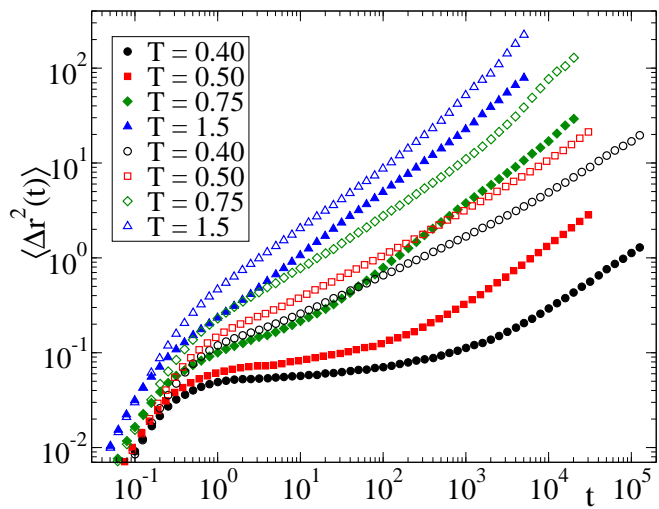}
\end{center}
\caption{Mean squared displacement at different temperatures
for both components. Filled and empty symbols correspond, respectively, to A- and B-chains.} 
\label{fig:msd}
\end{figure}

Fig. \ref{fig:msd} shows results for the mean squared displacement 
averaged over all the monomers, $\langle \Delta r^{2}_{\alpha}(t)\rangle$,
both for A- and B-chains. The introduction of monomer
size disparity, $\sigma_{\rm AA}/\sigma_{\rm BB} = 1.6$,
induces a large time scale separation between the two components,
for low concentration of the B-chains, by decreasing temperature.
Now we analyze relaxation features for the slow A-component.
Results for the fast B-component are analyzed in the next section. 

\begin{figure}
\begin{center}
\includegraphics[width=0.55\linewidth]{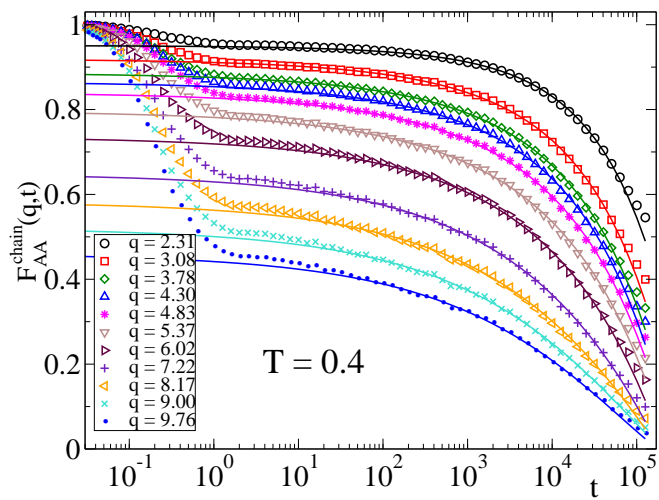}
\end{center}
\caption{Symbols: For different wave vectors, intrachain coherent correlator 
for A-A pairs, $F_{\rm AA}^{\rm chain}(q,t)$, at $T = 0.4$ .
Lines are fits to Eq. (\ref{eq:vonsch}) with an exponent $b = 0.30$.} 
\label{fig:fcohqtchainaa}
\end{figure}

\begin{figure}
\begin{center}
\includegraphics[width=0.55\linewidth]{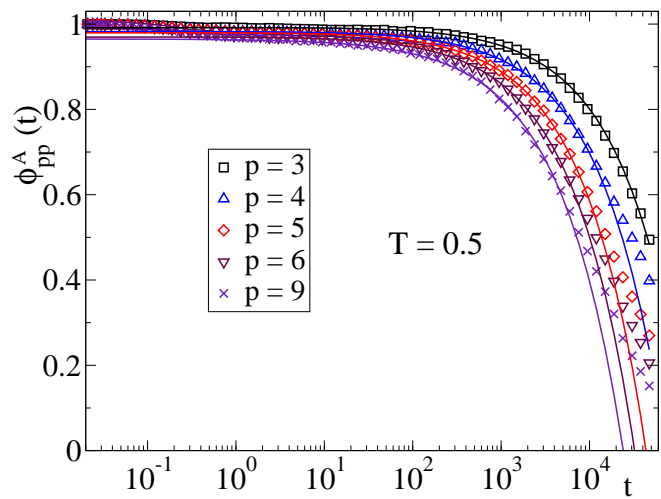}
\end{center}
\caption{Symbols: For different values of $p$, correlators of the Rouse $p$th-modes, $\phi_{pp}^{\rm A}(t)$,
of the A-chains. Lines are fits to Eq. (\ref{eq:vonsch}) with an exponent $b = 0.30$.
The temperature is $T = 0.5$.} 
\label{fig:rousea}
\end{figure}

\begin{figure}
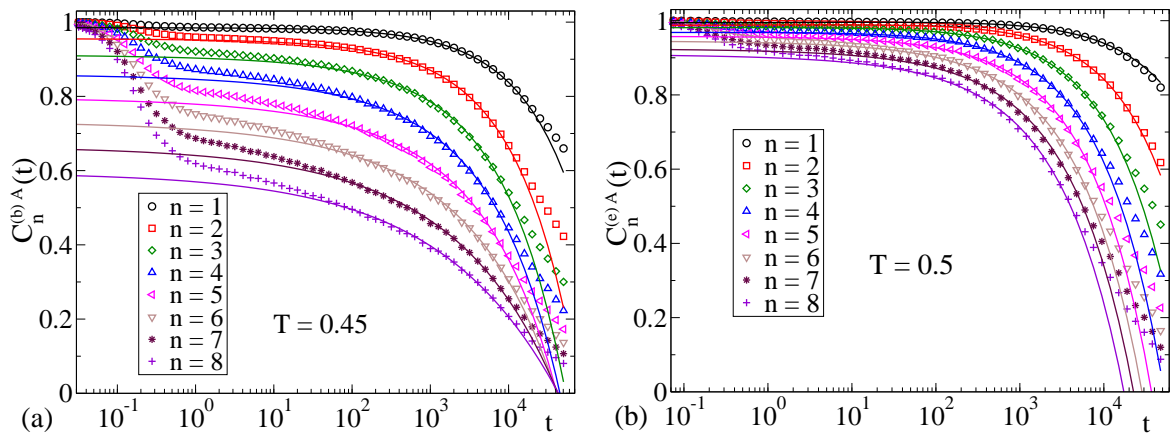

\begin{center}
\includegraphics[width=0.48\linewidth]{figure6a.eps}
\hspace{1 mm}
\includegraphics[width=0.48\linewidth]{figure6b.eps}
\end{center}
\caption{Panel (a): Symbols correspond, for different values of $n$, to angular correlators $C_n^{\rm (b) A}(t)$
of the bond vector of the A-chains. Lines are fits to Eq. (\ref{eq:vonsch}) with an exponent $b = 0.30$.
The temperature is $T = 0.45$. Panel (b): As panel (a) for angular correlators $C_n^{\rm (e) A}(t)$
of the end-to-end vector. The temperature is $T = 0.5$.} 
\label{fig:corbondeea}
\end{figure}

\begin{figure}
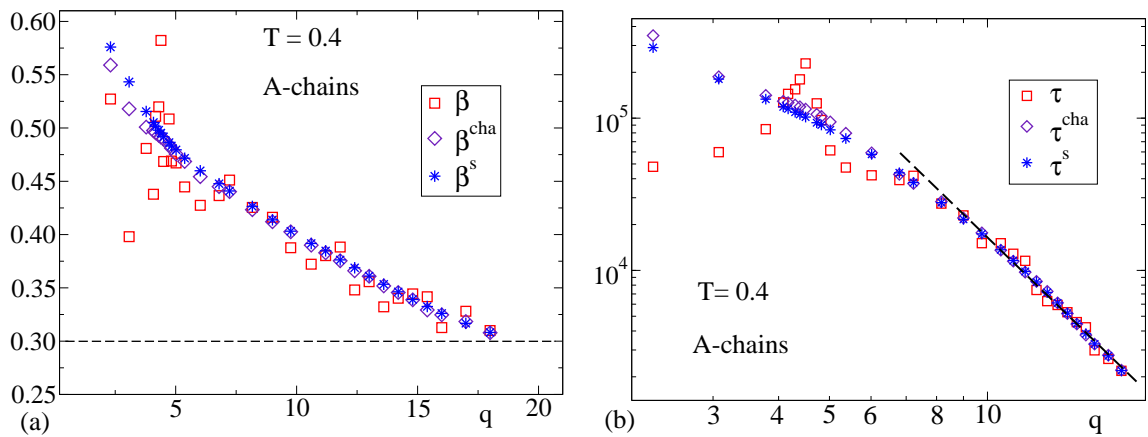

\begin{center}
\includegraphics[width=0.47\linewidth]{figure7a.eps}
\hspace{1 mm}
\includegraphics[width=0.47\linewidth]{figure7b.eps}
\end{center}
\caption{Panel (a): $q$-dependence of the stretching exponent  
of different correlators for the A-monomers (see text for notations) 
at temperature $T = 0.4$.
The dashed line indicates the large-$q$ limit $\beta (q) = b = 0.30$. 
Panel (b): As panel (a) for the corresponding KWW times (see text).
The dashed line corresponds to the power law $\propto q^{-1/b}$, with $b = 0.30$.} 
\label{fig:qtaubeta}
\end{figure}

\begin{figure}
\begin{center}
\includegraphics[width=0.55\linewidth]{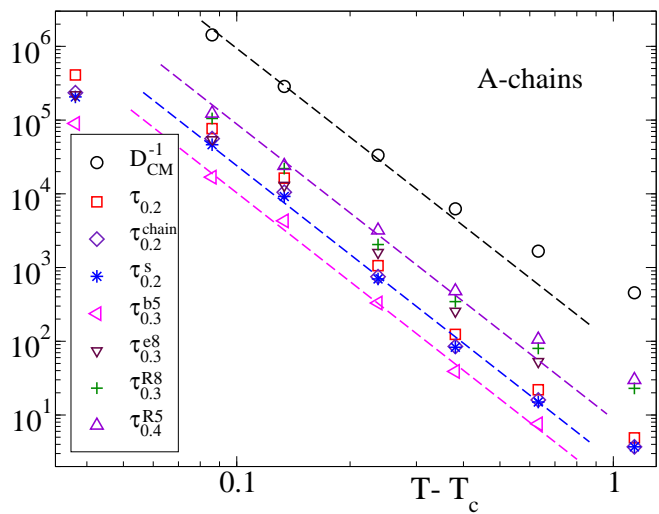}
\end{center}
\caption{Symbols: Inverse diffusivity (circles) and relaxation times of different correlators 
for the A-monomers (see text for notations). The wave vector for $\tau_{0.2}$,
$\tau_{0.2}^{\rm chain}$, and $\tau_{0.2}^{\rm s}$ is $q = 4.6$. 
The dashed lines are (from top to bottom) fits of
$D_{\rm CM}^{-1}$, $\tau_{0.4}^{\rm R5}$, $\tau_{0.2}^{\rm s}$,
and $\tau_{0.3}^{\rm b5}$ to the MCT power law $\propto (T-T_{\rm c})^{-\gamma}$, 
with $T_{\rm c}= 0.37$ and $\gamma = 4.0$.} 
\label{fig:mcttaua}
\end{figure}

Figs. \ref{fig:fcohqtchainaa}, \ref{fig:rousea}, \ref{fig:corbondeea},
\ref{fig:qtaubeta} and \ref{fig:mcttaua} show, for the A-chains, a consistent test
--- i.e., with a common set of exponents --- of MCT predictions for several dynamic correlators,
diffusivities, and relaxation times. 
Fig. \ref{fig:fcohqtchainaa} shows, for several wave vectors, results at $T = 0.4$ 
for the intrachain coherent correlator $F^{\rm chain}_{\rm AA}(q,t)$. The latter is computed as
$\left\langle \sum_{j,k} \exp\{i{\bf q}\cdot [{\bf r}_{\alpha,j}(t)-{\bf r}_{\alpha,k}(0)]\} \right\rangle / 
[N N_{\alpha} S_{\alpha\alpha}^{\rm chain}(q)]$, for any species $\alpha$. 
In this equation the sum only includes $j,k$ pairs belonging to a same $\alpha$-chain. 
Fig. \ref{fig:rousea} shows normalized correlators of the Rouse modes $\phi_{pp}^{\rm A}(t)$ at $T = 0.5$.
The latter are defined as 
$\Phi_{pp}^{\alpha}(t) = \langle {\bf X}^{\alpha}_p (t) \cdot 
{\bf X}^{\alpha}_p (0)\rangle /\langle [X^{\alpha}_p(0)]^2\rangle$, where the Rouse
normal modes \cite{teraoka,doiedwards}
of index $p = 0,1,...N-1$ are given by 
${\bf X}_p (t) = N^{-1}\sum_{j=1}^{N}{\bf r}_j (t) \cos [jp\pi/N]$.
Fig. \ref{fig:corbondeea}a displays, at $T = 0.45$, angular correlators
$C_n^{\rm (b) A}(t)$ for the bond vector, ${\bf b}(t)$,
between consecutive monomers. Such correlators are defined as $C_n^{\rm (b) A}(t) = P_{n}[\cos\theta(t)]$,
where $P_{n}$ is the Legendre polynomial of order $n$, 
and $\cos\theta(t) = \langle {\bf b}(t)\cdot {\bf b}(0) \rangle / \langle b^2 (0) \rangle$.
Angular correlators $C_n^{\rm (e) A}(t)$ for
the chain end-to-end vector, ${\bf e}(t)$, are defined in analogous way, with 
$\cos\theta(t) = \langle {\bf e}(t)\cdot {\bf e}(0) \rangle / \langle e^2 (0) \rangle$.
Data for $C_n^{\rm (e) A}(t)$ at $T = 0.5$ are given in Fig. \ref{fig:corbondeea}b.

Lines in Figs. \ref{fig:fcohqtchainaa}, \ref{fig:rousea}, and \ref{fig:corbondeea}
are fits of the $\alpha$-decay of the mentioned correlators to a power-law series expansion 
as Eq. (\ref{eq:vonsch}) with a common von Schweidler exponent $b = 0.30$. 
Only terms up to second order ($t^{2b}$) are included in the fit procedure
(in the following, references to this equation will be understood as limited to second order).
It must be stressed that the validity of Eq. (\ref{eq:vonsch}) for
the early-middle $\alpha$-decay must not be assessed by the length of the {\it vertical} interval of $\Phi (t)$
that it is able to cover. Indeed, if the relaxation time of the analyzed correlator
is much longer than the $\alpha$-time $\tau_{\alpha}$, 
the vertical interval described by (\ref{eq:vonsch}) will be rather small, as we will discuss below. 
The prefactors $h_{\Phi}$ and $h_{\Phi}^{(2)}$ in (\ref{eq:vonsch}), which yield the amplitude
of the decay, are generally in anti-phase with $\Phi^{\rm c}$ \cite{mctrev3,franosch,fuchspra,gotzevoigtmann} 
and are small for large values of the latter.
Hence, for correlators with high plateaux Eq. (\ref{eq:vonsch}) will only describe a small {\it vertical} interval
of the decay. On the contrary, validity of (\ref{eq:vonsch}) is given
by the extension of the time window (i.e., {\it horizontal} interval) that it is able to describe.
In the present case a good description of the simulation data is obtained
over three time decades for the lowest investigated temperature, 
a time window of validity which is typically achieved in simulations.
It must be noted that such a time window corresponds to a specific dynamic regime,
the early-middle stage of the structural $\alpha$-relaxation. However, relaxation of a given correlator
to a small value (e.g., $x =0.2$) can occur at a very different time scale $\tau^{\Phi}_{x}$.
This is the case of, e.g., low-index correlators of Rouse modes or chain end-to-end vectors.
The latter show a decay much slower than density-density correlators $F_{\rm AA}(q,t)$ 
at the maximum of $S_{\rm AA}(q)$ ($q = 4.5$), which properly probe the time scale $\tau_{\alpha}$
of the structural $\alpha$-relaxation for the A-chains.
For $F_{\rm AA}(q,t)$ we find $\tau_{0.2} = 1.7\times 10^{4}$ at $T = 0.5$, 
while relaxation times at the same temperature
for low indexes of $\Phi^{\rm A}_{pp}(t)$ and $C^{\rm (e) A}_n (t)$ are clearly much longer 
(see Figs. \ref{fig:rousea} and \ref{fig:corbondeea}b). 
Having said so, MCT establishes than asymptotic expansions as (\ref{eq:vonsch}) will be observed
for any dynamic correlator in the specific early-middle time window of the $\alpha$-relaxation,
the proccess here investigated.

Fig. \ref{fig:qtaubeta} shows a test of the MCT predictions $\beta (q \rightarrow \infty) = b$, 
and $\tau (q \rightarrow \infty) \propto q^{-1/b}$.
The stretching exponents $\beta$, $\beta^{\rm cha}$, and $\beta^{\rm s}$,
correspond respectively to the density-density, intrachain coherent and self-correlators of the A-monomers
at $T = 0.4$, and are obtained as fits of the decay from the plateau to a KWW function.
The corresponding KWW times are respectively denoted
as $\tau$, $\tau^{\rm cha}$, and $\tau^{\rm s}$.
The mentioned large-$q$ predictions for stretching exponents and KWW times are fulfilled with $b = 0.30$,
i.e., with the same value of the von Schweidler exponent used in the fits of the dynamic correlators
presented in Figs. \ref{fig:fcohqtchainaa}, \ref{fig:rousea}, and \ref{fig:corbondeea}.

The exponent $b = 0.30$ provides through Eqs. (\ref{eq:gamma}) and (\ref{eq:lambda}) the values
$\lambda = 0.90$, $a = 0.21$, and $\gamma = 4.0$. Now we test the validity of Eq. (\ref{eq:power}) with
this latter value of $\gamma$. Fig. \ref{fig:mcttaua} shows the temperature dependence
of the relaxation times $\tau^{\Phi}_x$ of several dynamic correlators $\Phi$
for the A-monomers. As mentioned above, these times are defined as those 
where the corresponding correlator decays to a value $x$.
Notations $\tau_{0.2}$, $\tau_{0.2}^{\rm chain}$, $\tau_{0.2}^{\rm s}$, $\tau_{0.3}^{\rm b5}$, $\tau_{0.3}^{\rm e8}$,
$\tau_{0.3}^{\rm R8}$, and $\tau_{0.4}^{\rm R5}$ correspond, respectively, to the correlators $F_{\rm AA}(q,t)$,
$F_{\rm AA}^{\rm chain}(q,t)$, $F_{\rm A}^{\rm s}(q,t)$, $C_5^{\rm (b) A}(t)$, $C_8^{\rm (e) A}(t)$,
$\phi_{88}^{\rm A}(t)$, and $\phi_{55}^{\rm A}(t)$. The wave vector for the first three correlators is $q = 4.6$,
an intermediate value between the main maxima of $S_{\rm AA}(q)$ and $S_{\rm AA}^{\rm chain}(q)$ (see Fig. \ref{fig:sq}).
The selected values of $x$ are well below the plateau
height of the correlator (see Figs. \ref{fig:fcohqtchainaa}, \ref{fig:rousea}, and \ref{fig:corbondeea}).
Also displayed is the inverse diffusivity, $D_{\rm CM}^{-1}$, of the center-of-mass of the A-chains,
which is defined as the long time limit of the ratio $6t/\langle [\Delta r_{\rm A}^{\rm CM}(t)]^2 \rangle$,
where $\langle [\Delta r_{\rm A}^{\rm CM}(t)]^2 \rangle$ is the corresponding mean squared displacement.

Dashed lines in Fig. \ref{fig:mcttaua} represent fits to the power law (\ref{eq:power}) by forcing
a common critical temperature $T_{\rm c}$ for all the data sets, with a fixed exponent $\gamma = 4.0$,
i.e., the value independently determined from the analysis previously presented 
in Figs. \ref{fig:fcohqtchainaa}, \ref{fig:rousea}, \ref{fig:corbondeea} and \ref{fig:qtaubeta}.
A value $T_{\rm c} = 0.37$ provides the best global fit with the mentioned constraint.
Interestingly, this value is much lower than the one obtained for the homopolymer state \cite{blendpaper}
at the same investigated packing fraction $\phi = 0.53$, $T_{\rm c} = 0.52$.
The latter value is obviously identical for the limits $x_{\rm B} = 0$ and $x_{\rm B} = 1$,
since the energy scale $\epsilon$ of the model is the same for all the pair interactions,
which only differ by the length scale $\sigma_{\alpha\beta}$ (see Section 2). 
Hence, blending at a fixed packing fraction stabilizes the ergodic phase as compared
to the homopolymer state, in analogy with the behaviour observed for colloidal binary mixtures
of similar size disparity \cite{gotzevoigtmann,williams,foffiprl}.

A good description of all the data sets is obtained over more than two decades
in relaxation time and inverse diffusivity.
As expected, due to the asymptotic character of Eq. (\ref{eq:power}), deviations from power-law
behaviour occur at high temperature. Such deviations are also present 
below some ill-defined temperature very close to $T_{\rm c}$. This feature is often observed
if one investigates dynamics at sufficiently low temperatures \cite{ashwin,gallo,flenner}
and is usually related with the presence of activated hopping events, which are not accounted for
within the ideal version of the MCT.

The analysis of simulation data of the slow A-component that has been presented 
in Figs. \ref{fig:fcohqtchainaa}, \ref{fig:rousea}, \ref{fig:corbondeea}, 
\ref{fig:qtaubeta}, and \ref{fig:mcttaua}
consists of a series of independent tests of several predictions of MCT with a common set of values
of the critical exponents. Therefore it provides a robust determination of such values,
and in particular of the exponent parameter $\lambda = 0.90$ from which the rest of the exponents
are derived through Eqs. (\ref{eq:gamma}) and (\ref{eq:lambda}). This value of $\lambda$ is unusually large,
as compared to those typical of one-component systems, as monodisperse 
hard spheres \cite{franosch} ($\lambda = 0.74$), 
simplified models of orthoterphenyl \cite{mossa} ($\lambda = 0.76$),
silica \cite{horbachsil} ($\lambda = 0.71$), water \cite{sciortinowat} ($\lambda = 0.78$),
or bead-spring homopolymers \cite{baschnagelrev} ($\lambda = 0.72$). In the following subsection 
we discuss the consequences of the large value of $\lambda$ here obtained.

\section{Dynamics of the fast component in the blend}

\begin{figure}
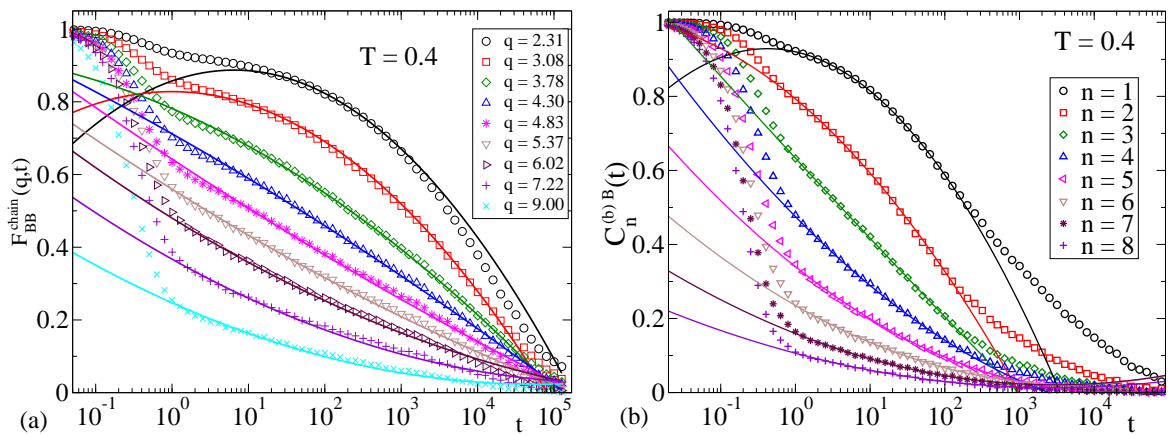

\begin{center}
\includegraphics[width=0.48\linewidth]{figure9a.eps}
\hspace{1 mm}
\includegraphics[width=0.48\linewidth]{figure9b.eps}
\end{center}
\caption{Panel (a): Symbols correspond, for different wave vectors, 
to intrachain coherent correlators 
for B-B pairs, $F_{\rm BB}^{\rm chain}(q,t)$, at $T = 0.4$.
Lines are fits to the  logarithmic expansion (\ref{eq:log}).
Panel (b): As panel (a) for the angular correlators $C_n^{\rm (b) B}(t)$
of the bond vector of the B-chains.} 
\label{fig:fcohqtchainbb}
\end{figure}

\begin{figure}
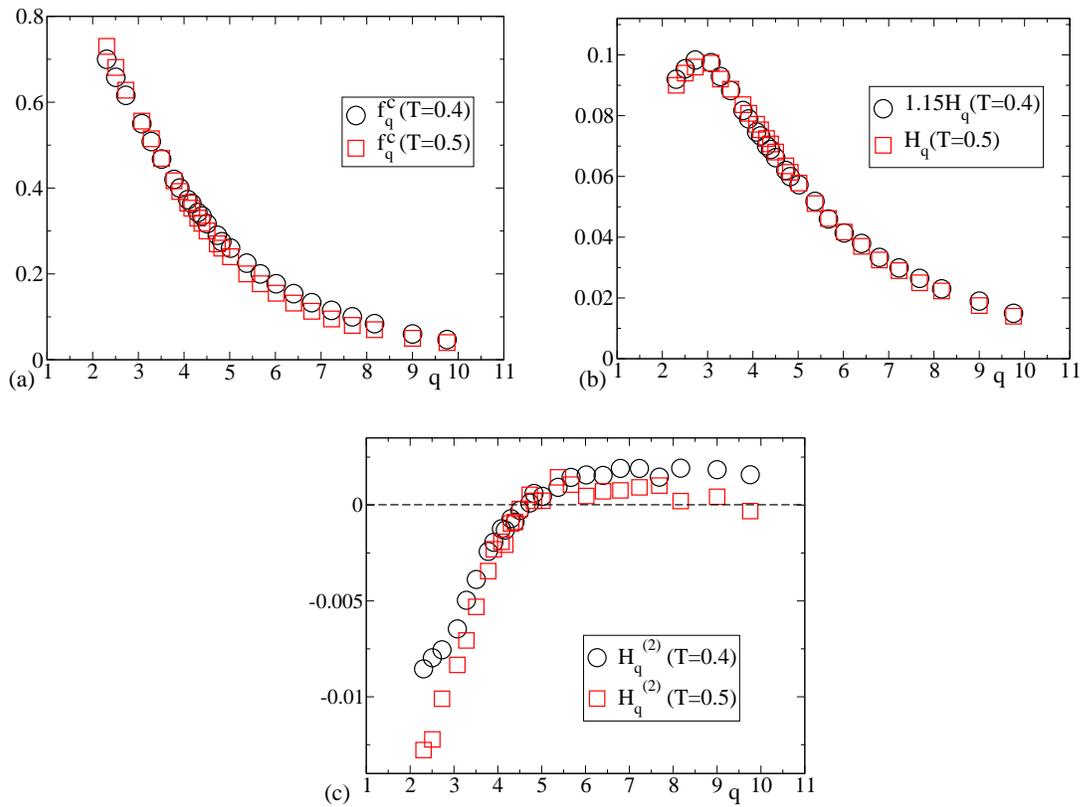

\begin{center}
\includegraphics[width=0.43\linewidth]{figure10a.eps}
\hspace{5 mm}
\includegraphics[width=0.43\linewidth]{figure10b.eps}
\newline
\newline
\includegraphics[width=0.43\linewidth]{figure10c.eps}
\end{center}
\caption{Symbols in panels (a), (b), and (c) correspond respectively to the values 
of the critical non-ergodicity parameter, $f^{\rm c}_{q}$, and the prefactors 
$H_q$ and $H_q^{(2)}$ in Eq. (\ref{eq:log}), for the intrachain coherent correlator
for B-B pairs, $F^{\rm chain}_{\rm BB}(q,t)$.
Temperatures are $T = 0.4$ (circles) and $T = 0.5$ (squares).} 
\label{fig:fqhqbb}
\end{figure}

Now we analyze the dynamics of the fast B-component.
Fig. \ref{fig:fcohqtchainbb}a shows simulation results for the intrachain coherent correlator,
$F_{\rm BB}^{\rm chain}(q,t)$, at temperature $T = 0.4$.
As previously reported in Ref. \cite{blendpaper} for the total density-density 
correlator $F_{\rm BB}(q,t)$, a concave-to-convex crossover is observed by varying the wave vector.
For intermediate values of the latter, a purely logarithmic decay occurs over more than three time decades.
Following a procedure analogous to that of Ref. \cite{blendpaper}, we have analyzed
the decay of  $F_{\rm BB}^{\rm chain}(q,t)$ in terms of a logarithmic expansion,
\begin{equation}
F_{\rm BB}^{\rm chain}(q,t) = f^{\rm c}_q -H_q \ln(t/\tau_{\sigma}) 
+ H_q^{(2)}\ln ^2 (t/\tau_{\sigma}) + O[\ln ^3 (t/\tau_{\sigma})],
\label{eq:log}
\end{equation}
with $\tau_{\sigma} \sim t_{\sigma}$,
instead of the von Schweidler series (\ref{eq:vonsch}) used for the A-monomers (we will discuss this point below). 
Within the framework of MCT, logarithmic expansions of dynamic correlators are associated
to the presence of a nearby higher-order transition \cite{schematic,dawson,sperl}. 
The latter is characterized by a value of the exponent parameter $\lambda = 1$, though
analogous predictions are expected for sufficiently large values $\lambda \rightarrow 1^{-}$
as the one here obtained, $\lambda = 0.90$ (see Section \ref{secslow}).
Eq. (\ref{eq:log}) provides a good description of the decay of correlators
displayed in Fig. \ref{fig:fcohqtchainbb}a. Analogous fits are shown in Fig. \ref{fig:fcohqtchainbb}b
for the orientational correlators $C_n^{\rm (b) B}(t)$ of the bond vector, 
evaluated for different values of $n$.
In this case the validity of the logarithmic expansion is observed, at the same temperature, 
over a shorter time interval.
Fig. \ref{fig:fqhqbb} shows the values of the coefficients
$f^{\rm c}_q$, $H_q$, and $H_q^{(2)}$ obtained from the corresponding fits of $F_{\rm BB}^{\rm chain}(q,t)$
at two different temperatures ($T = 0.4$ and $T = 0.5$).
The term $f^{\rm c}_q$ is the critical non-ergodicity parameter, which is associated to the transition point.
Therefore its values at different wave vectors must not depend on the state point at which 
they are obtained as fit parameters. This is confirmed by the numerical values displayed in Fig. \ref{fig:fqhqbb}a.
According to MCT, the prefactor $H_q$ is factorized as the product of two terms. One of them only depends
on the state point and the other one on the wave vector \cite{sperl}. Therefore the values of $H_q$ evaluated 
at different state points must obey scaling behaviour. This feature is also confirmed by data in Fig. \ref{fig:fqhqbb}b.
Also in agreement with MCT expectations \cite{sperl}, the obtained values of the second prefactor $H_q^{(2)}$
are smaller than $H_q$ and uncompatible with scaling behaviour (see Fig. \ref{fig:fqhqbb}c). 

\begin{figure}
\begin{center}
\includegraphics[width=0.55\linewidth]{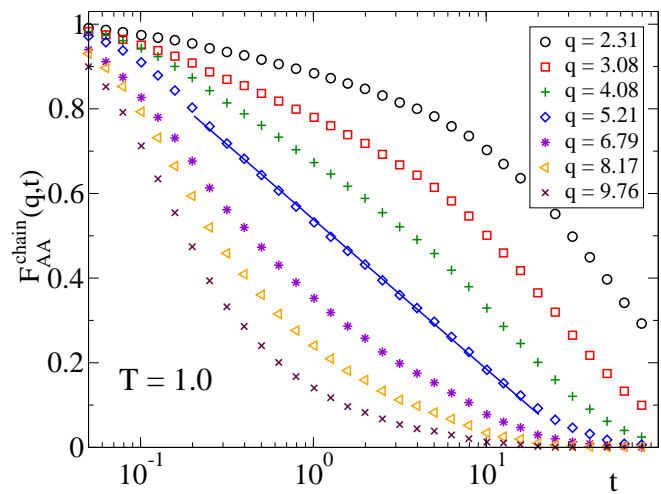}
\end{center}
\caption{Symbols: For different wave vectors, intrachain coherent correlator 
for A-A pairs, $F_{\rm AA}^{\rm chain}(q,t)$, at $T = 1.0$.
The straight line indicates logarithmic behaviour over two time decades.} 
\label{fig:fcohqtchainaaT1.0}
\end{figure}

Results presented in Figs. \ref{fig:fcohqtchainbb} and \ref{fig:fqhqbb} 
support the similar analysis performed in Ref. \cite{blendpaper} for 
density-density correlations of the B-monomers, $F_{\rm BB}(q,t)$. 
It must be stressed that the choice of Eq. (\ref{eq:log}) for describing relaxation of
correlators for B-monomers is not, in principle,
in contradiction with the description of the same correlators for the A-monomers 
in terms of the power-law series (\ref{eq:vonsch}). Both equations are {\it series expansions}
whose convergence depends on the analyzed region of the control parameter space.
For the case of higher-order transitions ($\lambda = 1$),
or more generally for transitions with $\lambda \rightarrow 1^{-}$, there are $q$-dependent paths
in the control parameter space where the series (\ref{eq:log}) is rapidly convergent. 
In particular for each wave vector there are optimal paths where $H_q^{(2)} = 0$. Along these paths
the corresponding correlator will exhibit a purely logarithmic decay \cite{sperl}. 
Moreover, by properly tuning the control parameters
or the wave vector, it is possible to change the signus of $H_q^{(2)}$ and, as a consequence, 
inducing a concave-to-convex crossover in the shape of the decay \cite{sperl}, 
as observed in Fig. \ref{fig:fcohqtchainbb}.
Since from the analysis of dynamic correlators for the A-monomers
we have determined a value $\lambda = 0.90$, it might be expected that such correlators will exhibit
such features at some state point. Indeed, they are observed at higher temperatures, 
as shown in Fig. \ref{fig:fcohqtchainaaT1.0} for $F_{\rm AA}^{\rm chain}(q,t)$ at $T = 1.0$.
The decay exhibits a clear concave-to-convex crossover by tuning the wave vector. 
Logarithmic relaxation covers two time decades for $q \approx 5.2$.

The fact that features associated to nearby higher-order MCT transitions
are observed for the A- and the B-component at very different temperatures 
must be commented. As mentioned above, the optimal paths in the control parameter space
for the observation of logarithmic relaxation are different for each correlator \cite{sperl}. 
The location of these paths is controlled by static correlations \cite{sperl},
which in the present case are very different for the A- and the B-monomers (see Fig. \ref{fig:sq}).
This difference might explain why anomalous relaxation for both types of monomers is observed at very
different temperatures. Unless one moves close to the optimal path in the control parameter space 
--- which usually involves a simultaneous variation of {\it several} control parameters \cite{schematic,sperl}--- 
logarithmic relaxation vanishes by decreasing temperature, and a standard two-step decay is recovered
(see Ref. \cite{zaccarelli1} for an illustrative example).
This seems to be the case of correlators for the A-monomers, which at low temperature are well described
by the von Schweidler power-law series (\ref{eq:vonsch}). A similar result has also been observed for mixtures
of large and small non-bonded particles \cite{mixturejcp}.
Still, a satisfactory answer to this point can only be obtained by solving the MCT equations for this system.

\begin{figure}
\begin{center}
\includegraphics[width=0.55\linewidth]{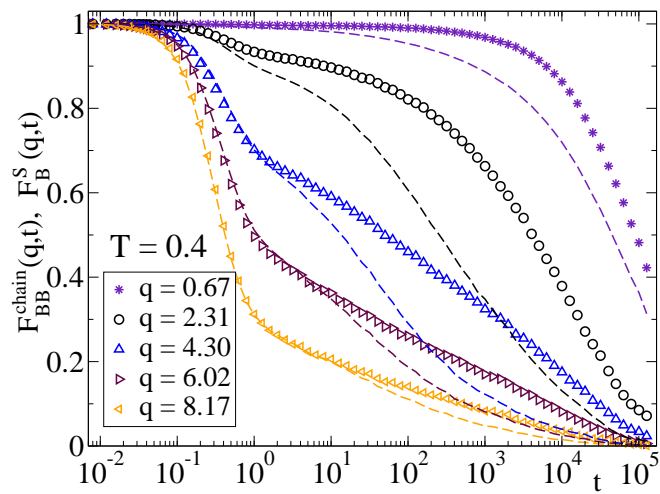}
\end{center}
\caption{Comparison, at $T = 0.4$, between the intrachain
coherent ($F^{\rm chain}_{\rm BB}(q,t)$, symbols) and 
self- ($F^{\rm s}_{\rm B}(q,t)$, dashed lines) correlators for the B-monomers. 
The wave vectors for the latters are, from top to bottom, 
the same ones as for the formers (see legend).} 
\label{fig:compchaincb}
\end{figure}

Fig.\ref{fig:compchaincb} displays for the B-chains at $T = 0.4$, results for 
the density self-correlators, $F^{\rm s}_{\rm B}(q,t)$.
A reliable fit (over more than one time decade)
of the corresponding decays to Eq. (\ref{eq:vonsch}) --- with any common exponent $b$ for all wave vectors --- 
or (\ref{eq:log}) was not possible. Similar tests were also unsuccesful
for correlators probing reorientations of chain end-to-end vectors, $C_n^{\rm (e) B}(t)$,
and relaxation of Rouse modes, $\Phi_{pp}^{\rm B}(t)$, of the B-chains. 
The reason for the apparent failure for the B-chains, or at least limited range
of validity, of the former equations for these correlators remains to be understood.
It might be that this feature is connected to a non-universal character 
of the asymptotic expansions (\ref{eq:vonsch}), (\ref{eq:log}) for binary mixtures with 
very different time scales for their respective density fluctuations, and that despite this
non-universality, MCT can still reproduce the behaviour of the mentioned correlators for the B-chains.
It might either be related to the presence of hopping events intervening in self-motions
of the B-monomers (see below).
As we argue in the following, there are results in the literature that support these possibilities. 

Numerical solutions of the MCT equations have been recently presented for sodium
silicate melts and compared with simulation results \cite{voigtmannhorbach}. 
In these systems the fast sodium atoms and the
slow silica matrix exhibit a strong time scale separation similar to that observed here
for the A- and B-chains \cite{voigtmannhorbach,horbachkob}. Though an analysis of density-density
correlators for the different atomic species, as well as of self-correlators for silicon and oxygen,
have provided a consistent test of MCT predictions with a common set of dynamic exponents
\cite{horbachkob}, self-correlators $F^{\rm s}(q,t)$ of the sodium atoms
do not show \cite{horbachkob}, as in the present case, a reliable time interval 
for apparent validity of Eqs. (\ref{eq:vonsch}) or (\ref{eq:log}).
Still, the corresponding numerical solutions of MCT equations reported 
in Ref. \cite{voigtmannhorbach} do reproduce 
the qualitative behaviour of $F^{\rm s}(q,t)$ for the sodium atoms. In particular,
MCT gives account for the unusual time scale separation between self- and collective
density correlators which is observed for the sodium atoms.  This feature is assigned
\cite{horbachkob,jund,greaves,lammert,meyer} to preferential diffusion along a long-living structure of channels
induced by the much slower relaxation of the silica matrix, which leads, for the alkali ions,
to a fast decay of self-correlations as compared to collective correlations.

\begin{figure}
\begin{center}
\includegraphics[width=0.46\linewidth]{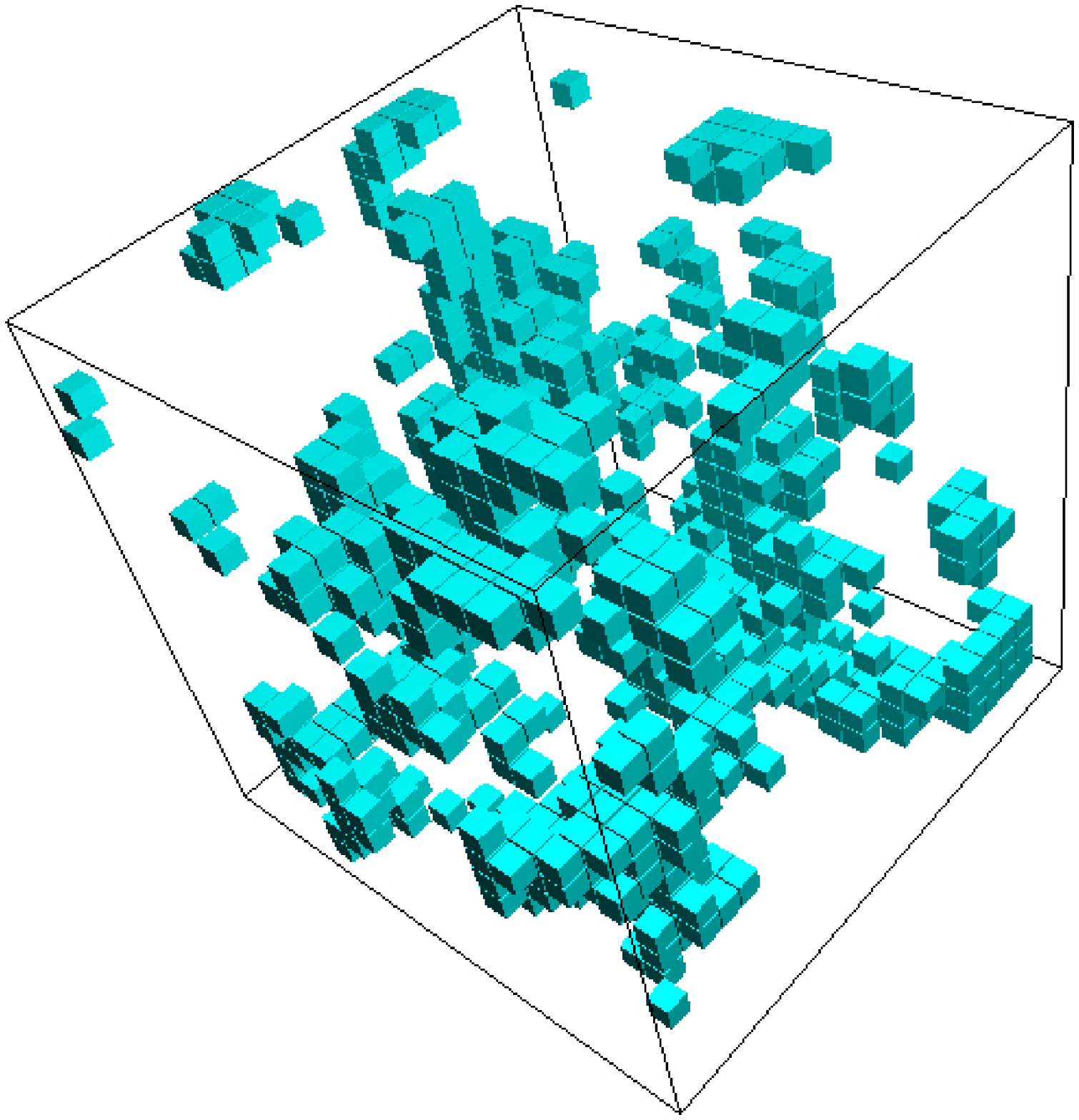}
\hspace{3 mm}
\includegraphics[width=0.49\linewidth]{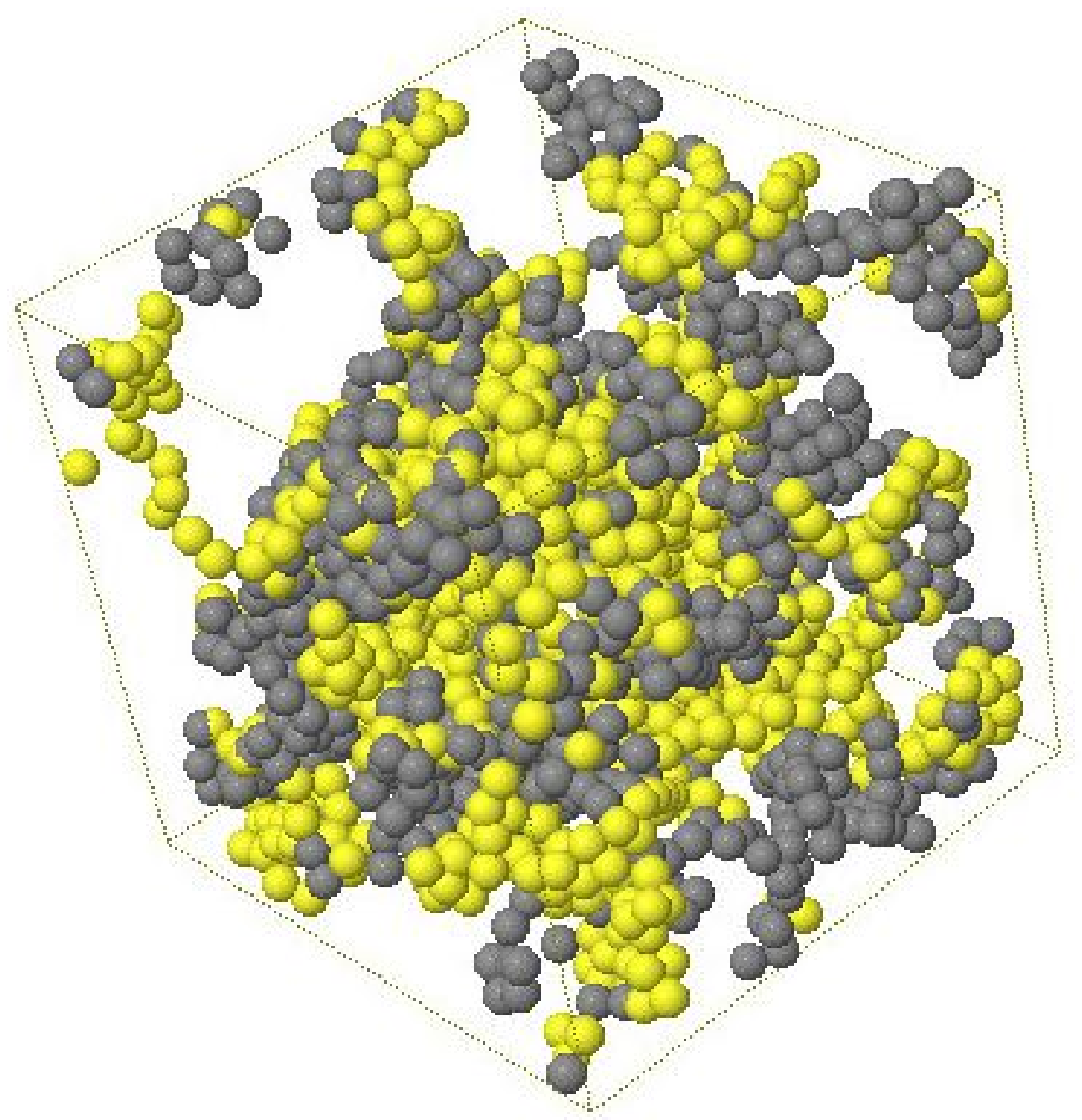}
\end{center}
\caption{Left side: Cubic boxes represent the $N_{\rm B}N$ subcells (of size $\approx \sigma_{\rm BB}^3$) 
in the simulation box which have been visited more times by the B-monomers during
a simulation time of $t = 5 \times 10^4$, for $T = 0.5$. Right side: Initial (dark spheres) 
and final (light spheres) configuration of the B-monomers for the latter simulation interval.
The same orientation of the simulation cell is used in both figures.}
\label{fig:channel0.3}
\end{figure}

Fig. \ref{fig:compchaincb} shows a comparison between self- and intrachain
coherent correlators for the B-chains at $T = 0.4$. 
Both correlators only converge to each other in the limit of large-$q$.
Since, due to the low concentration
of the B-component, intrachain coherent correlators for the B-chains exhibit only
small differences (not shown) with density-density correlators for {\it all} the B-B pairs,
the large time scale separation between $F^{\rm s}_{\rm B}(q,t)$  and  $F^{\rm cha}_{\rm BB}(q,t)$ 
presented in Fig. \ref{fig:compchaincb} is a feature analogous to
that above commented for alkali ions in silica matrices \cite{horbachkob}.
Indeed, following a procedure similar to that presented in Ref. \cite{jund} for the sodium atoms, we have determined
a similar structure of channels for preferential motion of the B-chains.
We have divided the simulation box in cubic subcells 
of size $\approx \sigma_{\rm BB}^3$ and computed, for a trajectory of the system,
the number of times each subcell is visited by a B-monomer. Fig. \ref{fig:channel0.3} displays,
at $T = 0.5$, the $N_{\rm B}N$ (a number equal to that of B-monomers)
most visited subcells for a simulation time $t = 5 \times 10^4$. 
The latter is much longer than the time for structural relaxation of the B-monomers at that temperature
As shown in Fig. \ref{fig:channel0.3}, the mentioned subcells are not randomly distributed
but form connected clusters, in analogy with results reported in \cite{jund} for sodium in silica matrices.

In Fig. \ref{fig:channel0.3} we also display the
initial and final configuration of the B-monomers for the  mentioned simulation interval $t = 5 \times 10^4$
used for the computation of the most visited subcells. As expected ($t$ is much longer than
the structural relaxation time for B-monomers) both configurations are fully decorrelated.
Therefore the mentioned channel structure is not a trivial consequence 
of the {\it static} correlations for the B-B pairs, which also
form a cluster structure (Fig. \ref{fig:configb}). It is instead induced by the time
scale separation of the {\it dynamic} correlations, which are much slower for the confining matrix
formed by the A-chains. The channel structure will only vanish when any region
of the simulation cell will be visited by the B-monomers with the same probability.
This can only occur at much longer times probing full structural relaxation of the A-component. 
A detailed static and dynamic characterization of this channel structure 
is beyond the scope of this article and will be presented elsewhere. 

Finally, it must be mentioned that the observed decoupling
between intrachain collective and self-correlators is exhibited only by the B-chains in the blend.
For the A-chains in the blend, as well as for the homopolymers,
we have observed only small differences for the latter correlators.
Decoupling between self- and collective intrachain dynamics is indeed a rather unusual feature, 
at odds with expectations from the standard Rouse model \cite{teraoka,doiedwards}. This observation in the
simple bead-spring blend here investigated is supported by recent neutron scattering experiments
on PEO/PMMA at low concentration of PEO \cite{niedzwiedz}.
Whether numerical solutions of MCT equations are also able, in analogy with the case of alkali ions in silica, 
to give account for this feature is an opened question.

\begin{figure}
\begin{center}
\includegraphics[width=0.52\linewidth]{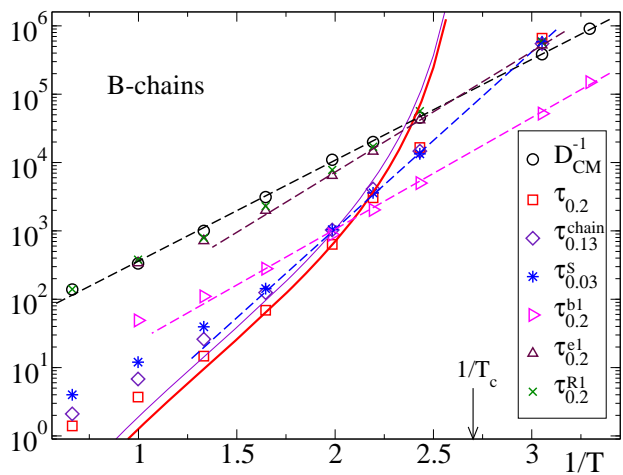}
\end{center}
\caption{Symbols: Inverse diffusivity (circles) and relaxation times of different correlators 
for the B-monomers (see text for notations). The wave vector for $\tau_{0.2}$,
$\tau_{0.13}^{\rm chain}$, and $\tau_{0.03}^{\rm s}$ is $q = 4.6$. 
The thick and thin solid lines are fits of, respectively, $\tau_{0.2}$ and $\tau^{\rm chain}_{0.13}$
to a MCT power law $\propto (T-T_{\rm c})^{-\gamma}$, with $T_{\rm c}= 0.37$ and $\gamma = 4.0$.
The arrow  indicates the inverse value of $T_{\rm c}$. The dashed lines are (from top to bottom)
fits of $D_{\rm CM}^{-1}$, $\tau_{0.2}^{\rm e1}$, $\tau_{0.2}^{\rm b1}$, 
and $\tau_{0.03}^{\rm s}$ to Arrhenius behavior, $\propto \exp(E/T)$. The activation energies 
are respectively $E = 3.4$, 4.1, 3.8, and 5.9.} 
\label{fig:taub}
\end{figure}

Fig. \ref{fig:taub} shows the temperature dependence of the
relaxation times for several dynamic correlators probing relaxation of the B-component.
Notations $\tau_{0.2}$, $\tau_{0.13}^{\rm chain}$, 
$\tau_{0.03}^{\rm s}$, $\tau_{0.2}^{\rm b1}$, $\tau_{0.2}^{\rm e1}$,
and $\tau_{0.2}^{\rm R1}$ correspond, respectively, to the correlators $F_{\rm BB}(q,t)$,
$F_{\rm BB}^{\rm chain}(q,t)$, $F_{\rm B}^{\rm s}(q,t)$, $C_1^{\rm (b) B}(t)$, $C_1^{\rm (e) B}(t)$,
and $\phi_{11}^{\rm B}(t)$. The wave vector for the first three correlators is $q = 4.6$.
Also included is the inverse diffusivity of the center-of-mass for the B-chains.
The set of data shown in Fig. \ref{fig:taub} exhibits a behaviour rather different from
similar quantities for the A-chains displayed in Fig. \ref{fig:mcttaua}.
Only relaxation times for collective density correlations, $F_{\rm BB} (q,t)$ and $F_{\rm BB}^{\rm chain}(q,t)$,
show qualitative agreement with the MCT power-law $\propto (T-0.37)^{-4.0}$ derived from data of the A-monomers.
As expected, deviations occur at temperatures very close to $T_{\rm c}$. The rest of the quantities
displayed in Fig. \ref{fig:taub} are uncompatible with power-law behaviour. They show instead an apparent
Arrhenius dependence, $\propto \exp(E/T)$, from moderate to the lowest investigated temperature. 
This feature is demostrated in Fig. \ref{fig:taub} by the linear behaviour observed by representing 
data in logarithmic scale vs. (linear) $1/T$.
The obtained activation energies $E$ vary between 3.4 for the center-of-mass diffusivity and 5.9 
for the relaxation time of density self-correlations. The observed Arrhenius behaviour suggest that strong
hopping events intervene in the structural relaxation of the B-chains, similarly to observations
for alkalin ions in silica \cite{horbachgeol}. These events seem to affect more strongly
to self- than to collective density
correlations, for which a power law behaviour can be observed over two decades in relaxation time
for temperatures above $T_{\rm c}$. It remains to be understood whether such hopping 
events --- which are not included in the ideal version of MCT --- are related to the mentioned reduction 
of the range of validity of Eqs. (\ref{eq:vonsch}) or (\ref{eq:log}) for the corresponding correlators.
It is worth mentioning that the latter possibility might be the case for
sodium atoms in silica. Numerical solutions of the MCT equations reported in Ref. \cite{voigtmannhorbach},
though reproducing the observed qualitative behaviour, understimate the strength of the decay
exhibited in simulations for self-correlators $F^{\rm s}(q,t)$ of sodium. Hence, the presence of hopping
events presumably accelarates relaxation as compared to theoretical predictions.

Finally, it is worthy of remark that the observed Arrhenius-like temperature
dependence for $\tau_{0.2}^{\rm b1}$ and $\tau^{\rm s}_{0.03}$ is consistent with experimental
observations, for the fast component,
in real polymer blends with large dynamic asymmetry by,
respectively, dielectric spectroscopy \cite{dielec1,dielec2,dielec4} and neutron scattering \cite{tyagi},
which probe relaxation times of similar dynamic correlators. Arrhenius behaviour for self-dynamics 
is also observed for the case
of alkali ions in silica \cite{horbachgeol} or for water reorientation in polymer matrices \cite{cerveny}.
This common Arrhenius-like behaviour in very different systems suggests a universal feature 
for low concentrations of fast molecules in slow host media with interconnected voids.

\section{Conclusions}

We have presented a computational investigation on the structural relaxation of 
a simple bead-spring model for polymer blends with large dynamic asymmetry.
We have computed a large set of dynamic correlators probing relaxation of density fluctuations, 
Rouse modes, and reorientation of bond and chain end-to-end vectors.
Results have been discussed within the framework of the Mode Coupling Theory (MCT)
for the glass transition. A robust test of MCT predictions has been achieved through a description 
of the different analyzed correlators with a common set of dynamic exponents,
though for some correlators probing dynamics of the fast component MCT asymptotic laws are apparently not observed. 
The observation of Arrhenius-like behaviour suggests that this breakdown 
might be associated to strong hopping events intervening in relaxation of the fast component.

An unusually large value of the exponent parameter $\lambda$ has been obtained, close to the upper limit
($\lambda = 1$) characteristic of higher-order MCT transitions. According to MCT predictions,
the anomalous relaxation features observed in the present system, as logarithmic decays 
or concave-to-convex crossovers in density correlators, might be associated to that
underlying higher-order scenario. An investigation of the case of extreme dilution,
where each individual chain of the fast component is sorrounded only by chains of the slow component
(and where an asymptotic dynamic limit is expected \cite{nmr2}), would be computationaly expensive. 
Still, we expect that a qualitatively similar scenario of anomalous relaxation will be observed. 
Since chain connectivity will always guarantee the presence of neighbouring monomers
of the same species for a given monomer of the fast component, coexistence of bulk-like caging
and confinement for the fast component would be present even at extreme dilution, 
inducing the higher-order scenario.
On the other hand, a progressive increase of the concentration of the fast component will
reduce the time scale separation (i.e., the dynamic asymmetry) between the two components,
and confinement effects will finally vanish. 
In that situation a standard MCT relaxation scenario (as observed for
the homopolymer case) will be recovered.  
The large collection of results here presented
might motivate theoretical work on structural relaxation in polymer blends with large dynamic asymmetry,
and in particular, numerical solutions of the MCT equations to confirm the suggested higher-order scenario.

\section{Acknowledgments}

We acknowledge financial support from the projects NMP3-CT-2004-502235 (SoftComp), 
MAT2004-01017 (Spain), and 206.215-13568/2001 (GV-UPV/EHU Spain). A.J.M acknowledges
support from DIPC (Spain).



\section*{References}

\begin{thebibliography}{}


\bibitem{kumar} Kant R, Kumar S K and Colby R H 2003 {\it Macromolecules} {\bf 36} 10087 
\bibitem{lodge} Lodge T P and McLeish T C B 2000 {\it Macromolecules}  {\bf 33} 5278
\bibitem{kumar2} Kumar S K, Shenogin S and Colby R 2007 {\it Macromolecules} {\bf 40} 5759
\bibitem{blendsag} Cangialosi D, Schwartz G A, Alegr\'{\i}a A 
and Colmenero J 2005 {\it J. Chem. Phys.} {\bf 123} 144908 \\
Schwartz G A, Cangialosi D, Alegr\'{\i}a A and Colmenero J 2006 {\it J. Chem. Phys.} {\bf 124} 154904
\bibitem{nmr1} Lartigue C, Guillermo A and Cohen-Addad J P 1997 {\it J. Polym. Sci. B: Polym. Phys.} {\bf 35} 1095
\bibitem{nmr2} Lutz T R, He Y Y, Ediger M D, Cao H H, Lin G X and Jones A A 2003 {\it Macromolecules} {\bf 36} 1724
\bibitem{dielec1} Lorthioir C, Alegr\'{\i}a A and Colmenero J 2003 {\it Phys. Rev. E} {\bf 68} 031805  
\bibitem{dielec2} Sy J W and Mijovic J 2000 {\it Macromolecules} {\bf 33} 933
\bibitem{dielec3} Roland C M, McGrath K J and Casalini R 2006 {\it Macromolecules} {\bf 39} 3581
\bibitem{dielec4} Schwartz G A, Colmenero J and Alegr\'{\i}a A 2007 {\it Macromolecules} {\bf 40} 3246
\bibitem{dielec5} Taneko H, Kobayashi M and Aikawa T 2006 {\it Macromolecules} {\bf 39} 2183
\bibitem{genix} Genix A C, Arbe A, Alvarez F, Colmenero J, Willner L and Richter D 2005 {\it Phys. Rev. E} {\bf 72} 031808
\bibitem{tyagi} Tyagi M, Arbe A, Colmenero J, Frick B and Stewart J R 2006 {\it Macromolecules} {\bf 39} 3007
\bibitem{blendpaper} Moreno A J and Colmenero J 2006 {\it J. Chem. Phys.} {\bf 124} 184906 
\bibitem{mctrev1} G\"{o}tze W 1991 {\it Liquids, Freezing and Glass Transition, Les Houches 1989} 
ed J P Hansen, D Levesque and J Zinn-Justin (Amsterdam: North-Holland) p 287
\bibitem{mctrev2} G\"{o}tze W and Sj\"{o}gren L 1992 {\it Rep. Prog. Phys.} {\bf 55} 241
\bibitem{schematic} G\"{o}tze W and Haussmann R 1988  {\it Z. Phys. B: Condens. Matter} {\bf 72} 403 \\
G\"{o}tze W and Sperl M 2002 {\it Phys. Rev. E} {\bf 66} 011405  
\bibitem{dawson} Dawson K, Foffi G, Fuchs M, G\"{o}tze W, Sciortino F, Sperl M, Tartaglia P,
Voigtmann T and Zaccarelli E 2000 {\it Phys. Rev. E} {\bf 63} 011401 
\bibitem{sperl} Sperl M 2003  {\it Phys. Rev. E} {\bf 68} 031405 
\bibitem{prlsimA4} Sciortino F, Tartaglia P and Zaccarelli E 2003 {\it Phys. Rev. Lett.} {\bf 91} 268301 
\bibitem{zaccarelli1} Zaccarelli E, Foffi G, Dawson K A, Buldyrev S V,
Sciortino F and Tartaglia P 2002 {\it Phys. Rev. E} {\bf 66} 041402 
\bibitem{mixturepre} Moreno A J and Colmenero J 2006 {\it Phys. Rev. E} {\bf 74} 021409
\bibitem{mixturejcp} Moreno A J and Colmenero J 2006 {\it J. Chem. Phys.} {\bf 125} 164507
\bibitem{mayer} Mayer C 2007 Ph. D. Thesis (Universit\"{a}t D\"{u}sseldorf)
\bibitem{bingemann} Bingemann D, Wirth N, Gmeiner J and R\"{o}ssler E A 2007 {\it Macromolecules} {\bf 40} 5379
\bibitem{krakoviack} Krakoviack V 2007 {\it Phys. Rev. E} {\bf 75} 031503
\bibitem{grest} Grest G S and Kremer K 1986 {\it Phys. Rev. A} {\bf 33} R3628
\bibitem{koband} Kob W and Andersen H C 1995 {\it Phys. Rev. E} {\bf 51} 4626 \\ 
Kob W and Andersen H C 1995 {\it Phys. Rev. E} {\bf 52} 4134
\bibitem{bennemann} Bennemann C, Paul W, Binder K and Dunweg B 1998 {\it Phys. Rev. E} {\bf 57} 843 \\ 
Bennemann C, Baschnagel J and Paul P 1999  {\it Eur. Phys. J. B}  {\bf 10} 323
\bibitem{frenkel} Frenkel D and Smit B 1996 {\it Understanding Molecular Simulation} (San Diego: Academic Press)
\bibitem{aichele} Aichele M and Baschnagel J  2001 {\it Eur. Phys. J. E}  {\bf 5} 229 \\
Aichele M and Baschnagel J  2001 {\it Eur. Phys. J. E}  {\bf 5} 245
\bibitem{teraoka} Teraoka I  2002  {\it Polymer Solutions} (New York: John Wiley \& Sons)
\bibitem{doiedwards} Doi M and Edwards S F 1986 {\it The Theory of Polymer Dynamics} (Oxford: Oxford University Press)
\bibitem{mctrev3} G\"{o}tze W  1999 {\it J. Phys.: Condens. Matter} {\bf 11} A1
\bibitem{das} Das S P 2004 {\it Rev. Mod. Phys.}  {\bf 76} 785
\bibitem{franosch} Franosch T, Fuchs M, G\"{o}tze W, Mayr M R and Singh A P 1997 {\it Phys. Rev. E} {\bf 55} 7153 
\bibitem{fuchs} Fuchs M 1994 {\it J. Non-Cryst. Solids} {\bf 172} 241 
\bibitem{fuchspra} Fuchs M, Hofacker I and Latz A 1992 {\it Phys. Rev. A} {\bf 45} 898
\bibitem{mossa} Mossa S, Di Leonardo R, Ruocco G and Sampoli M 2000 {\it Phys. Rev. E} {\bf 62} 612
\bibitem{puertaspre} Puertas A M, Fuchs M and Cates M E  2003  {\it Phys. Rev. E} {\bf 67} 031406 
\bibitem{pisacolme} Colmenero J, Narros A, Alvarez F, Arbe A and Moreno A J  2007 
{\it J. Phys.: Condens. Matter} {\bf 19} 205127
\bibitem{gotzevoigtmann} G\"{o}tze W and Voigtmann T  2003  {\it Phys. Rev. E} {\bf 67} 021502  
\bibitem{williams} Williams S R and van Megen W  2001 {\it Phys. Rev. E}  {\bf 64} 041502
\bibitem{foffiprl} Foffi G, G\"{o}tze W, Sciortino F, Tartaglia P and Voigtmann T   2003
{\it Phys. Rev. Lett.} {\bf 91} 085701 
\bibitem{ashwin} Ashwin S S and Sastry S  2003 {\it J. Phys.: Condens. Matter} {\bf 15} S1253 
\bibitem{gallo} Gallo P, Pellarin R and Rovere M  2003 {\it Phys. Rev. E} {\bf 67} 041202
\bibitem{flenner} Flenner E and Szamel G  2005 {\it Phys. Rev. E} {\bf 72} 011205 
\bibitem{horbachsil} Horbach J and Kob W 1999 {\it Phys. Rev. B} {\bf 60} 3169 \\ 
Horbach J and Kob W 2001 {\it Phys. Rev. E} {\bf 64} 041503 
\bibitem{sciortinowat} Sciortino F, Fabbian L, Chen S H and Tartaglia P 1997 {\it Phys. Rev. E} {\bf 56} 5397
\bibitem{baschnagelrev} Baschnagel J and Varnik F 2005 {\it J. Phys.: Condens. Matter} {\bf 17} R851
\bibitem{voigtmannhorbach} Voigtmann T and Horbach J 2006  {\it Europhys. Lett.} {\bf 74} 459
\bibitem{horbachkob} Horbach J and Kob W 2002 {\it J. Phys.: Condens. Matter} {\bf 14} 9237 \\
Horbach J, Kob W and Binder K 2002 {\it Phys. Rev. Lett.} {\bf 88} 125502
\bibitem{jund} Jund P, Kob W and Jullien R  2001  {\it Phys. Rev. B} {\bf 64} 134303
\bibitem{greaves} Greaves G N  1985 {\it J. Non-Cryst. Solids} {\bf 71} 203
\bibitem{lammert} Lammert H, Kunow M and Heuer A  2003 {\it Phys. Rev. Lett.} {\bf 90} 215901 
\bibitem{meyer} Meyer A, Horbach J, Kob W, Kargl F and Schober H  2004 {\it Phys. Rev. Lett.} {\bf 93} 027801 \\ 
Kargl F, Meyer A, Koza M M and Schober H  2006 {\it Phys. Rev. B} {\bf 74} 014304 
\bibitem{niedzwiedz} Niedzwiedz K, Wischnewski A, Monkenbusch M, Richter D, Genix A C, Arbe A,
Colmenero J, Strauch M and Straube E 2007 {\it Phys. Rev. Lett.} {\bf 98} 168301
\bibitem{horbachgeol} Horbach J, Kob W, Binder K  2001 {\it Chem. Geol.} {\bf 174} 87
\bibitem{cerveny} Cerveny S, Schwartz G A, Bergman R, Swenson J  2004 {\it Phys. Rev. Lett.} {\bf 93} 245702 \\
Cerveny S, Schwartz G A, Alegr\'{\i}a A, Bergman R, Swenson J  2006 {\it J. Chem. Phys.} {\bf 124} 194501 









 

\end{thebibliography}
\end{document}